\documentclass[preprint,11pt]{elsarticle}
\usepackage{graphicx}
\usepackage{bm}
\usepackage{amssymb}
\usepackage{wasysym}
\usepackage{mathrsfs}
\usepackage{epstopdf}
\usepackage{subfigure}
\usepackage{setspace}
\journal{International Journal of Modern Physics B}

\begin{document}
\begin{frontmatter}
\title{Non-relativistic fermionic energy gap in the non-abelian gauge systems}
\author{Bo-Jie Huang$^\dag$, Chyh-Hong Chern$^*$\footnote{chchern@ntu.edu.tw}}
\address{$^\dag$Institute of Atomic and Molecular Sciences, Academia Sinica, No. 1, Roosevelt Rd., Sec. 4, Taipei, 10617, Taiwan}
\address{$^*$Department of Physics, National Taiwan University, Taipei 10617, Taiwan}

%\date{\today}
\begin{abstract}
We demonstrate that the non-relativistic fermions open the energy gap when the SU(N) gauge bosons, mediating the interaction between fermions, acquire the mass.  Surprisingly, even though there is the SU(N) gauge symmetry, there is always one fermionic energy gap which is not degenerate to the rest of the $N-1$ fermions for $N \ge 3$ in the fundamental representation.
\end{abstract}


\end{frontmatter}

%\preprint{\sf Version 1 (\today)}

%\begin{figure}[htbp]
%\begin{center}
%\includegraphics[scale=0.5]{vertex.pdf}
%\caption{The vertex of the electron-electron interaction.}
%\label{vertex}
%\end{center}
%\end{figure}

\section{Introduction}
The energy gap formation is an ubiquitous phenomena in condensed matter systems.  When the band structure appears in the one-particle Hamiltonian with a periodic potential, the band gap is the region in the spectrum where there is no density of states.   On the other hand, the repulsion interaction generates the energy gap in the fractional quantum Hall systems.  Generally speaking, systems with an energy gap are more stable against perturbations.

The systems with the energy gap are, however, not good nurturing cradles for the superconductivity, which arises in the systems with Fermi surface (gapless).  Because of the instability of the interaction with the phonons, the electrons pair up and condense to the superconducting state.  However, there are some classes of superconductors which were obtained by doping the antiferromagnetic insulators with mobile carriers, for example high transition temperature superconductors in the cooper-based transition metal oxides (cuprates)~\cite{bednorz, wu}.  By the chemical doping, the systems enter the phase where the energy gap structure is anisotropic in the momentum space, before becoming the superconductors~\cite{shen}.  The enigmatic gap phase has agonised condensed matter community for three decades.

Recently, one of us (Chern) developed a weak-coupling theory based on the Hubbard model for the gap formation in cuprates~\cite{chc}.  Introducing the spin Berry's phase as the gauge interaction~\cite{nagaosa}, the Hubbard model in two dimensions can be formulated in the renormalizable theory in the continuous limit.  Considering the antiferromagnetic fluctuation additionally, the gauge field acquires the mass via the St\"uckelberg mechanism.  The 2+1 dimensional Lagrangian density is given by
\begin{eqnarray}
\mathcal{L}= \sum_{\sigma}{\psi}^{\dagger}_{\sigma}(x)(i{\partial}_{0}-ga_{0}){\psi}_{\sigma}(x)-\frac{1}{2m}[(-\frac{\vec {\nabla}}{i}-g\vec a){\psi}^{\dagger}_{\sigma}(x)][(\frac{\vec {\nabla}}{i}-g\vec a){\psi}_{\sigma}(x)]   \nonumber \\
-\frac{1}{4}f_{\mu\nu}f^{\mu\nu}+M_{0}(D_{0}{\phi}(x))^{\dagger}(D_{0}{\phi}(x))-M_{1}(\vec D{\phi}(x))^{\dagger}{\cdot}(\vec D{\phi}(x)), \label{u1}
\end{eqnarray}
where $\psi_\sigma$ are the electrons, $\vec{a}$ are the gauge fields, $g$ is the gauge coupling, $\phi$ is the antiferomagnetic fluctuation, $D_\mu$ are the covariant derivatives, and $M_0$ and $M_1$ are the mass parameters.  The antiferromagnetic fluctuation is parameterised by a complex phase field $\frac{1}{q}e^{i\sigma(x)}$, where $q$ is the coupling between the gauge field and the antiferromagnetic fluctuation. In two dimensions, the $\phi$ field takes place an infinite order phase transition at the finite temperature, so called the Berezinski-Kosterlitz-Thouless transition~\cite{fisher, thouless}.   Combining with the gauge fields, the $\phi$ field becomes the longitudinal mode of the gauge fields.   As the transition of the mass acquisition takes place, the electronic energy structure opens a gap without breaking the translational and the time reversal symmetry.

The gap formation is not the patent for cuprates but has found in many other strongly-correlated electron systems, for example the iron pnictides and the heavy fermion systems~\cite{thompson, hosono}. Unlike the cuprates, the iron pnictides and the heavy fermion materials are the multi-band systems.  It inspires us to generalise the current U(1) scheme to the SU($N$) cases, where the multiple $N$-flavours of electrons can be considered.  Furthermore, while the St\"uckelberg mechanism works in the U(1) case, we generalise the mass acquisition scheme to the Higgs mechanism.  Restricting ourself to the simplest fundamental representation for both electrons and the Higgs, we found that there is always one flavour of the electrons which is not degenerate to the other for $N \ge 3$.  This robust behaviour can be understood by the group theory.

In this paper, the sections are organised as the following.  In the second section, the SU(2) case will be discussed.  In the third section, the results of the SU($N$) cases are provided.  The last section is the discuss and the conclusion.

\section{The SU(2) case}
\subsection{The SU(2) Lagrangian}
For a system with multi-flavours of electrons that are degenerate to each other, we can possibly consider the SU(2) gauge symmetry.  For simplicity, we consider the electrons to be in the SU(2) fundamental representation.  The U(1) Lagrangian in Eq.~(\ref{u1}) can be generalised to the SU(2) form, 
\begin{eqnarray}
\mathcal{L}_{0}= {\psi}^{\dagger}(x)(i{\partial}_{0}-ga_{0}){\psi}(x)-\frac{1}{2m}[(\frac{\vec {\nabla}}{i}-g\vec a){\psi}(x)]^{\dagger}[(\frac{\vec {\nabla}}{i}-g\vec a){\psi}(x)]   \nonumber \\
-\frac{1}{4}f_{\mu\nu}f^{\mu\nu}+M_{0}^{2}(D_{0}{\phi}(x))^{\dagger}(D_{0}{\phi}(x))-M_{1}^{2}(\vec D{\phi}(x))^{\dagger}{\cdot}(\vec D{\phi}(x)), \label{lagrangian}
\end{eqnarray}
where $\psi(x)=(\psi_1(x), \psi_2(x))^{T}$, $D_0=i{\partial}_{0}-g'a_{0}$, $\vec{D}_0=-i\vec{\nabla}-g'\vec{a}$, $g$ and $g'$ are the gauge couplings for the electrons and the Higgs boson respectively, and $a_0$, $\vec{a}$, and $f_{\mu\nu}$ are matrix-valued,
\begin{eqnarray}
a_{i} =a^{c}_{i}\frac{\sigma^{c}}{2},  \ a_{0} = {a}^{c}_{0}\frac{\sigma^{c}}{2},\ f_{\mu\nu} = f_{\mu\nu}^{c}\frac{\sigma^{c}}{2},
\end{eqnarray}
where $\sigma^c$ are the Pauli spin matrices.  The Higgs field can be stabilised by the following terms
\begin{eqnarray}
\mathcal{L}'=\frac{{\mu}^{2}}{2}{\phi}^{2}-\frac{\lambda}{4}{\phi}^{4} \label{higgs},
\end{eqnarray}
where ${\mu}$ is the Higgs mass and ${\lambda}$ is self-interaction parameter.  The total Lagrangian density is given by $\mathcal{L}=\mathcal{L}_{0}+\mathcal{L}'$.

The mass generation of the SU(2) gauge bosons via the Higgs mechanism is a textbook story.  For example, the mass acquisition of the gauge boson is related to the group representation of the Higgs field.   In the fundamental representation, three gauge bosons acquire the equal mass, and in the adjoint representation, only two gauge bosons obtain the mass.  On the other hand, different from the high-energy physics, the condensed matter community cares more about the length scale.  The gauge bosons of zero mass produce a long-ranged interaction, and the ones of finite mass produce a short-ranged interaction.  In the condensed matter systems, the long-ranged interaction is often screened and becomes short-ranged.  In the systems with the gauge symmetry, it corresponds to the gauge bosons of finite mass~\cite{chc}.

\subsection{Energy gap formation}

As the gauge bosons acquire the mass, the short-ranged interaction modifies the electronic specturm, opening a gap-like structure in the non-relativistic band structure~\cite{chc}.  In the condensed matter language, the notion of the energy gap is different from the mass, which is determined by the curvature of the dispersion relation.  The nature of the phase transition to the gap phase is, however, different from the U(1) case.  In the Higgs mechanism given by Eq.~(\ref{higgs}), it favors a second-order phase transition.  In the real materials, it may take place at the finite temperature, if the two dimensionality of the space is only an approximation.

\begin{figure}[htbp]
\includegraphics[scale=0.6]{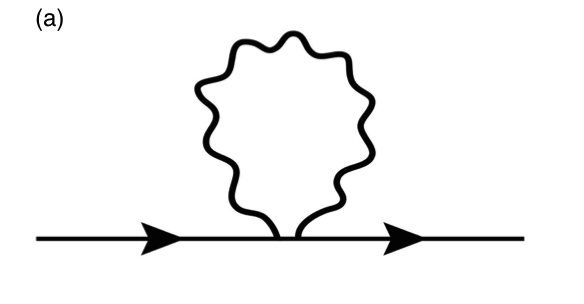}
\includegraphics[scale=0.3]{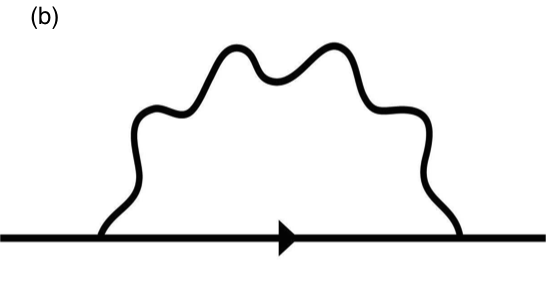}
\caption{The one-loop diagrams contributing to the electron self-energy.}
\label{diagram}
\end{figure}

Similar to the U(1) case, we compute the energy gap using the single-particle Green's function.  The leading diagrams contributing to the self-energy term $\Sigma(\omega,p)$ are given in the Fig.~(\ref{diagram}). 
In the fundamental representation of the Higgs mechanism, the electronic gap, the energy at the bottom of the band, is
\begin{eqnarray}
\Delta_{f}=\frac{3g^{2}}{4{\pi}m}\frac{g'{\nu}}{4}(M_{1}+\frac{M^{2}_{0}}{3M_{1}})
\end{eqnarray}
for both flavors of the electrons.  Although the diagram in Fig.~(\ref{diagram}b) modifies the dispersion relation, it does not contribute to the gap generation.  On the other hand, in the adjoint representation of the Higgs mechanism, it becomes
\begin{eqnarray}
\Delta_{a}=\frac{\sqrt{2}g^{2}}{4{\pi}m}\frac{g'{\nu}}{4}(M_{1}+\frac{M^{2}_{0}}{3M_{1}})
\end{eqnarray}
for all flavors of the electrons.

The SU(2) theory may be realized in the condensed matter system with the non-abelian holonomy~\cite{chc} and the magnetism.  The non-abelian holonomy plays the role of the SU(2) gauge fields.  On the other hand, the ferromagnetic or the antiferromagnetic fluctuations may serve as the Higgs field.  If  the non-abelian holonomy is in the particle-hole channel of the degrees of freedom, for example the spin Berry's phase, it may be able to couple to the (anti)-ferromagnetic fluctuation and manifests the effect of the electronic gap generation.

\section{The SU(N) case}

The mechanism of the non-relativistic gap generation can be generalized to the SU($N$) case.  The formalism of the SU($N$) Lagrangian is the same as the ones in Eq.~(\ref{lagrangian}) and Eq.~(\ref{higgs}).  In addition, the electrons are considered in the SU($N$) fundamental representation, namely $\psi(x)=(\psi_1(x), ...,\psi_N(x))^{T}$.  If the Higgs field is also considered in the fundamental representation, the mass spectrum of the $N^{2}-1$ gauge bosons can be given as the following.
\begin{eqnarray}
&&m_{i}=0, \ \ \  \ \ \ \ \ \ \ \ \ \ \ \ \ \ \ \ \ \ \ 1\leqslant i\leqslant N^{2}-2N \nonumber\\
&&m_{i}=\frac{g'{\nu}}{2}, \ \ \ \ \ \ \ \ \ \  \ \ \ \ \ \ \ \ N^{2}-2N+1\leqslant i\leqslant N^{2}-2 \nonumber\\
&&m_{i}=\frac{g'{\nu}}{2}\sqrt{\frac{2(N-1)}{N}}, \ \ \ i=N^{2}-1 \label{higgsN},
\end{eqnarray}
where $N^2-2N$ gauge bosons remain massless, and the rest of them become massive.  Among the massive gauge bosons, there is always one boson acquiring different mass.  The self energy of the electrons is also computed using the diagram in Fig.~(\ref{diagram}).  We obtain
\begin{eqnarray}
&&\Delta_1=\frac{g^{2}}{4{\pi}m}\frac{g'{\nu}}{4}(M_{1}+\frac{M^{2}_{0}}{3M_{1}})\times (2+\sqrt{\frac{8}{N^{3}(N-1)}}),\ \ \ 1\leqslant i\leqslant N-1\nonumber \\
&&\Delta_2=\frac{g^{2}}{4{\pi}m}\frac{g'{\nu}}{4}(M_{1}+\frac{M^{2}_{0}}{3M_{1}})\times ((2N-2)+\sqrt{\frac{8(N-1)^{3}}{N^{3}}}),\ \ \ i=N \label{gapN}
\end{eqnarray}
 For $N=2$, we reproduce the results of the SU(2) case. Different from the SU(2) case, however, there is always one flavor of the electron that is not degenerate to the rest of the $N-1$ electrons.  This robust structure may be considered as the signature of the SU($N$) gauge symmetry for $N\ge 3$.
 
The current results can be understood by the group theory.   Before the symmetry breaking of the Higgs field, the theory is SU($N$) symmetric.  In the fundamental representation, there are $2N$ degrees of freedom in the $N$ multiplet of the Higgs field.  After the spontaneous symmetry breaking, there are $2N-1$ Goldstone modes which combine with the gauge bosons and become the longitudinal modes of the massive bosons.  Consequently, in the $N^2-1$ gauge bosons, there are $N^2-2N$ boson remaining massless as shown in Eq.~(\ref{higgsN}).  Interestingly, the remaining $N^2-2N$ bosons preserve the SU($N-1$) symmetry.  After the symmetry breaking, the remnant symmetry becomes SU($N-1$).  Therefore, spectrum of the $N$ electrons splits into $(N-1)+1$, reflecting the SU($N-1$) symmetry.

\section{Conclusion}
The nonrelativistic gap formation is generalized from the U(1) gauge symmetry with the St\"uckelberg mechanism to the SU($N$) gauge symmetry with the Higgs mechanism.  In the U(1) case, the phase transition is the Berezinskii-Kosterlitz-Thouless-like transition at the finite temperature in the 2+1 dimensional spacetime.  Namely, there is no significant signature of the phase transition.  On the other hand, in the SU($N$) case, the gap spectrum of the $N$-plet of the electrons splits into $(N-1)+1$, as the consequence of the remnant SU($N-1$) symmetry.  The SU($N$) theory may be applicable to the system with non-abelian holonomy.

\section{Acknowledgement}
We are grateful for the stimulated discussions with Chong-Der Hu and Pei-Ming Ho.  This work is supported by Ministry of Science and Technology of Taiwan under the grant: MOST 103-2112-M-002-014-MY3 and by Na- tional Taiwan University under the grant: 103R7831 and 104R7831.

%\bibliography{psdgap}
\end{document}